\documentclass{article}

\usepackage[letterpaper,margin=1in]{geometry}

\usepackage[utf8]{inputenc}
\usepackage[T1]{fontenc}
\usepackage{lmodern}
\usepackage{hyperref}
\usepackage{url}
\usepackage{booktabs}
\usepackage{amsfonts}
\usepackage{amsmath}
\usepackage{nicefrac}
\usepackage{microtype}
\usepackage[numbers,compress]{natbib}
\usepackage{placeins}
\usepackage{xcolor}
\usepackage{graphicx}
\usepackage{multirow}
\usepackage{booktabs}
\usepackage{authblk}

\title{Cross-Facility LLM Pre-training on HPC:\\ Elastic Aggregation, Data Leasing, and Queue-Aware Placement}

\author[1]{Zarè Palanciyan}
\author[1]{Thomas~van~Osch}
\author[1]{Douwe~van~der~Wal}
\author[2]{Olivera Kotevska}
\author[1]{Tim Kok}

\affil[1]{SURF, Amsterdam, The Netherlands}
\affil[2]{Oak Ridge National Laboratory, Oak Ridge, TN, USA}
\date{}

\begin{document}

\maketitle

\begin{abstract}
Academic compute is fragmented: allocations are granted per facility, and facilities differ in accelerators and software stacks, schedule jobs independently, and share neither a network nor a filesystem. We present a system that pools such allocations to pre-train a single language model across three supercomputers on two continents, up to 7{,}400\,km apart: Snellius (NVIDIA H100), LUMI and Frontier (both AMD MI250X). It combines (i) DiLoCo-style two-loop training with an elastic, token-weighted Nesterov outer step for which zero, one or many live sites are all normal states; (ii) DARL, a data-leasing protocol whose heartbeat-backed leases guarantee that, within an epoch, no sample is trained twice or lost under crashes, late joins and work stealing; and (iii) queue-aware placement driven by unprivileged \texttt{sbatch -{}-test-only} probes. Training Qwen3-0.6B on C4 for 20{,}000 optimiser steps, the three-site run reaches a held-out perplexity of 34.7, against 28.2 for a centralised baseline. Per-round overhead (weight exchange and checkpointing) stays near 110\,s regardless of the number of local steps $H$, so its share of wall-clock time falls from 32\% at $H{=}100$ to 5.7\% at $H{=}1{,}000$ and 3.1\% at $H{=}2{,}000$. In a 23.8\,h three-site run with four site departures, 1.2\% of granted data blocks were reclaimed and none was duplicated or lost. In an idealised queue-model projection, queue-aware placement shortens time-to-target by 18--43\% compared with waiting for all sites to be allocated. Cross-site pre-training is thus operational rather than competitive: it turns fragmented allocations into one training run at a measured cost.

\end{abstract}

\section{Introduction}\label{sec:intro}
Training frontier-grade foundation models in an acceptable time span requires more capacity than a single supercomputer in the public research sector provides. The individual allocations that high-performance computing (HPC) systems currently award are not enough to train a language model. Researchers must therefore aggregate capacity across national, regional and leadership-class facilities. That capacity is \emph{fragmented}, with allocation divided over different facilities, each with different hardware, a separate queue, and its own expiry (e.g., the 19 AI Factories and 13 AI Factory Antennas established by the European Union\citep{aifactories}). Academic groups must combine those pieces into one training run and not a series of unrelated jobs. This is the practical obstacle that researchers must overcome to train models at the scale of industry.

We therefore ask what an effective distributed training system looks like in a real deployment on public research infrastructure. Such a system must solve three problems simultaneously. The first is \emph{training logistics}: how sites divide the corpus and exchange weights when they share neither a filesystem nor a network. The second is \emph{metascheduling}: how to coordinate independent batch jobs at different facilities, when none can start on demand. The third is \emph{infrastructure}: every system brings its own accelerators, software stack, connectivity rules, and site policy. Leave any one unsolved, and the run does not start, does not finish, or does not converge.




The setup for this experiment is training a single 0.6B-parameter language model on disjoint shards of a dataset without repetition distributed across LUMI \citep{lumi, lumidocs}, Frontier \cite{frontier} and Snellius \citep{snellius}. Through the novel distributed allocation and remote leasing (DARL) leasing protocol, the run is flexible in late joins and scheduled churns. We also report a comparison against centralised training. Our contributions are as follows:

\textbf{Elastic aggregation.} A Flower-based~\citep{flower} outer loop applies a token-weighted Nesterov step~\citep{diloco,fedmom}. For this loop, 0, 1 or $N$ live sites are all normal states. A solo round takes the full outer step. A generation check rejects stale contributions. A joiner adopts the current global model and schedule. 

\textbf{DARL.} A dynamic data-leasing protocol. A coordinator leases disjoint blocks of the pre-tokenised corpus under heartbeat TTL contracts. The protocol guarantees that within an epoch no sample is trained twice and that no sample is lost, even as sites join and leave (\S\ref{sec:darl}).


\textbf{Live queue-aware planning.} Scheduled \texttt{sbatch --test-only} probes and \texttt{sacct} wall-time statistics feed a planner. The planner simulates the run round-by-round. It then chooses site membership, job shape and accumulation (\S\ref{sec:sched}).

\section{Related work}\label{sec:related}

\textbf{Low-communication and federated optimisation.} Local stochastic gradient descent (SGD) and its descendants~\citep{localsgd,postlocalsgd} established that replicas may take many local steps between synchronisations. DiLoCo~\citep{diloco} scaled the recipe to language models with an inner AdamW loop and an outer Nesterov step on averaged deltas. OpenDiLoCo~\citep{opendiloco} and INTELLECT-1~\citep{intellect1} then demonstrated geo-distributed runs at increasing scale. The aggregation itself descends from FedAvg~\citep{fedavg} and its server-side-momentum variants~\citep{fedavgm,fedopt,fedmom}. Photon~\citep{photon} reported that federated pre-training \emph{beat} a centralised baseline. At matched tokens, however, its federated configuration takes roughly $8\times$ more optimiser steps, with local batches with higher local learning rates than centralised. The comparison with a federation with a different optimisation regime is therefore improper. Our campaign matches tokens \emph{and} steps \emph{and} the combined per-step batch, which removes that mechanism by design. The two studies answer different questions.

\textbf{Geo-distributed systems and queue-aware placement.} Other work treats cross-site training as a systems and scheduling problem. Measurements of training on spot VMs across several continents show that the latency between continents and the ratio of compute to communication limit how well it scales, and that data-transfer fees can exceed the cost of the VMs~\citep{isenko2024clouds}. An analytical model shows that data parallelism across cloud regions slows down sharply, while pipeline parallelism keeps its throughput but costs more in transfer fees~\citep{strati2024crossregion}. DT-FM schedules training over decentralised GPUs joined by slow links~\citep{yuan2022dtfm}. Sailor chooses how many GPUs of each type to use in each cloud zone and re-plans when their availability changes~\citep{strati2025sailor}. None of these works models the batch queue through which HPC systems run every job.


Fairshare and backfill set start times, and walltime caps bound how long any replica can live. A job that ends is not preempted but simply finished. The relevant literature is therefore queue-wait and runtime prediction from scheduler history~\citep{smith1999runtime,brown2022queuewait,vercellino2023queueing}, which targets the placement of independent jobs within a single facility. Only recently has that signal reached training itself. xFFL orchestrates federated runs across facilities through a workflow manager~\citep{colonnelli2024xffl}. FedQueue predicts per-facility admission delays and absorbs them inside the aggregation rule, buffering late arrivals and down-weighting stale updates~\citep{li2026fedqueue}.

A field-tested LLM pre-training setup with elastic aggregation, dynamic data-leasing protocol, queue-aware distributed job allocation separates current work from existing literature.

\section{System}
\subsection{Two-loop training and elastic aggregation}\label{sec:fedmom}
Following DiLoCo, each site $i$ runs $H{=}100$ inner AdamW steps from the current global model $w^{(t)}$, then uploads its delta $\Delta_i^{(t)}=\theta_i^{(t)}-w^{(t)}$ together with the token count it consumed. With token shares $p_i$, the server applies an outer Nesterov step
\begin{equation}
v^{(t+1)} = w^{(t)} + \eta \textstyle\sum_i p_i \Delta_i^{(t)}, \qquad
w^{(t+1)} = v^{(t+1)} + \beta\,\bigl(v^{(t+1)} - v^{(t)}\bigr),
\label{eq:fedmom}
\end{equation}
with $\eta{=}0.7$, $\beta{=}0.9$~\citep{diloco}; our implementation is pinned in tests against \texttt{torch.optim.SGD} with Nesterov momentum to $2\times10^{-6}$. Setting $\beta{=}0,\eta{=}1$ recovers exact FedAvg. Three changes to Eq.~\ref{eq:fedmom} make the outer optimisation loop elastic, so sites can join or leave at any round. First, a round with a single live contributor takes the full step ($\eta{=}1$), instead of discarding $1-\eta$ of its update. Second, every delta records the merge round it was derived from, and the server rejects stale generations, a requeued job that trained from an old global model cannot poison the average. Third, the authoritative global step advances by the \emph{maximum} steps any contributor took. Weights cross the WAN as bfloat16 (1.32\,GiB per message for this model), inline in the gRPC message below the 2\,GiB cap and streamed tensor-by-tensor through the blob store above it, a small HTTP PUT/GET service on the same VM. For sites reachable only via an HTTP forward proxy that does not permit long-lived gRPC streams, the same round protocol runs exclusively over short HTTP requests.

\subsection{Distributed Allocation and Remote Leasing (DARL)}\label{sec:darl}
Clusters must not train on overlapping samples, and no sample may be lost when a node fails. A window trained twice in an epoch is counted twice in the token-weighted average of Eq.~\ref{eq:fedmom}, tilting the merged model toward whichever site duplicated it; a window dropped when a site vanishes shrinks the epoch with no record that it did. Heterogeneous sites also drain data at different rates and their availability is asynchronous, so any static partition is wasteful. DARL is a dynamic data-leasing protocol: a central coordinator holds the only authoritative map of which data blocks are unassigned, leased, or committed, and clusters lease contiguous blocks under heartbeat-backed time-to-live (TTL) contracts.

Every site holds an identical pre-tokenised dataset $D$ of $N$ sample windows. The coordinator partitions $D$ into $M=\lceil N/K\rceil$ blocks of $K$ consecutive samples; $K$ trades statistical mixing against I/O efficiency, since one block is one contiguous read. Because corpora such as C4 are not shuffled on disk, assigning block ranges naively would give each site a different topical slice of the corpus and induce drift between the site models, so DARL shuffles \emph{at block level} with a deterministic global permutation $\pi_s$ of the $M$ blocks, generated from a shared seed $s$ by the Durstenfeld variant of Fisher-Yates~\citep{durstenfeld1964algorithm}. A digest of $(N,K,s)$ is checked at registration, such that two sites cannot silently join with different block spaces. Lease granularity matches the DiLoCo inner phase, which consumes $H\!\cdot\!B\!\cdot\!W\!\cdot\!A$ samples ($H$ inner steps, per-rank batch $B$, $W$ ranks, $A$ accumulation steps): $M_{\text{phase}} = \max\bigl(1, \lceil H B W A / K \rceil\bigr)$ blocks per lease.

Leases expire on a TTL of $\alpha\,\bar t_{\text{step}} + \beta_{\text{RoundTripTime}}$ (with $\alpha{=}2.5$) refreshed by heartbeats, so a crashed site's untouched data returns to the pool automatically; a drained site steals the untouched tail of the largest outstanding lease; and grants are proportionally sized to each site's measured commit rate, so fast sites are neither starved nor hoarding. Exactly-once coverage is an ordering guarantee: a site checkpoints \emph{before} it commits blocks, so its weights and its data-consumption record can only fail together.

\subsection{Scheduling}\label{sec:sched}
Placement across facilities is a question no batch scheduler answers: each site schedules only its own machine, and prior cross-facility systems (\S\ref{sec:related}) hide admission delay rather than choose where to submit. Choosing requires queue-time estimates alongside each site's walltime cap and per-GPU throughput, and obtaining them is the hard part on production HPC, where most sites expose no scheduler API and do not permit cronjobs. All three of our sites run Slurm, so a single unprivileged mechanism suffices.

\paragraph{Observing queue times.}\label{sec:observe} For each site $c$ we define a set of \emph{job shapes} $\mathcal{S}_c$; a shape fixes a partition, node count, GPUs per node and walltime. Every $\langle\Delta\rangle$ the collector runs \texttt{sbatch -{}-test-only} per shape, placing the job against the live queue without enqueuing it. Slurm assumes every queued job runs to its requested walltime and is therefore systematically pessimistic; we discount each estimate by $\langle\gamma\rangle$, derived from realized walltime usage in a trailing \texttt{sacct} window.

\paragraph{Planning submissions.}\label{sec:plan}
Work is expressed in \emph{data units}, an arbitrary quantum of which one outer round consumes $\langle R\rangle$; each site carries a per-GPU rate from $\langle$historical job logs$\rangle$. A waterfill \citep{boyd2004convex} assigns outstanding units to the soonest-available site to minimise predicted round completion, splitting proportionally to throughput when several are free and capping each job at the site's walltime limit and optionally adding reserve time to compensate for other sites starting later. Parameters: a minimum units per job, the discount $\langle\gamma\rangle$, reserve $\langle\rho\rangle$, and a maximum probe age beyond which a site is excluded. The plan is recomputed every $\langle\Delta\rangle$, so newly free capacity is folded in without disturbing running jobs. Misprediction costs calendar time only: a job that starts late leases no shards (\S\ref{sec:darl}), so no work is lost or duplicated and the reserve $\langle\rho\rangle$ can dampen the impact.

\section{Results}\label{sec:results}
Training is done with torchtitan~\citep{torchtitan} on an Qwen3-0.6B model~\citep{qwen3}, trained on a pre-tokenised C4 dataset~\citep{c4}. Federation is managed by Flower \citep{flower}. Exact model details can be found in the Appendix \ref{app:setup}.

\begin{figure}[!h]
\centering
\includegraphics[width=\linewidth]{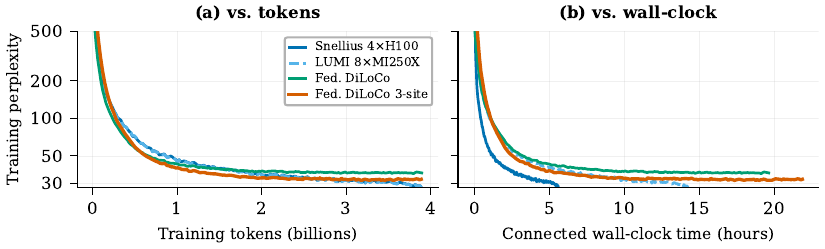}
\caption{Training perplexity against training tokens (a) and against connected wall-clock time (b). Both centralised runs use a batch size of 96. \emph{Fed.\ DiLoCo} is the two-site run (LUMI and Snellius). \emph{Fed.\ DiLoCo 3-site} is an elastic run over Frontier, LUMI and Snellius. LUMI and Frontier train together and then dropout, and Snellius continues the run on its own.}
\label{fig:quality}
\end{figure}


\textbf{Quality and time (Fig.~\ref{fig:quality}).} At matched tokens and matched steps, centralised training wins on quality and on wall-clock time. It reaches a perplexity of 28.2 in 5\,h\,41\,m on Snellius, against 37.5 in 19\,h\,48\,m for two-site DiLoCo. Two-site DiLoCo reaches an effective 55k tok/s, against 192k tok/s for centralised training. First, a higher peak learning rate is a large improvement \emph{without} federation (28.2\,$\rightarrow$\,24.5). Second, this outcome is the designed consequence of matching steps and per-step batch. Photon's federated advantage at matched tokens came from clients that took ${\sim}8\times$ more steps, and those steps higher local learning rate~\citep{photon}. Our controls remove that mechanism. 

\begin{figure}[!h]
\centering
\includegraphics[width=\linewidth]{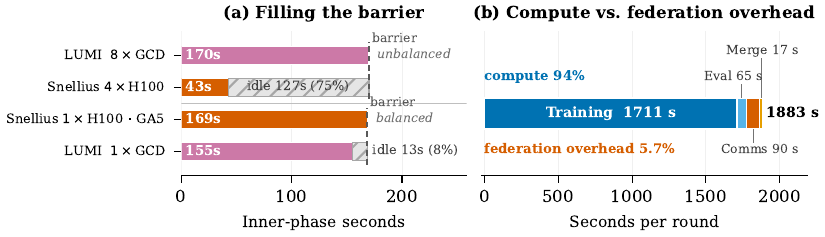}
\caption{(a) The unbalanced, $H{=}100$, at  fast site finishes 127\,s early and idles 75\% of the phase. With calibrated accumulation ($a_{\text{Snellius}}{=}5$) the sites land 13\,s apart, 8\% idle. (b) At a $H{=}1{,}000$, 1{,}881\,s round run 91\% of the compute is the same work a centralised run would also do. }
\label{fig:barrier}
\end{figure}

\textbf{Distributed overhead.}
Figure \ref{fig:barrier} shows that having equal global batch size on heterogenous hardware can stall all GPUs but the slowest. Filling the barrier with a larger global batch size on faster hardware, for example, increasing the gradient accumulation steps, can minimize this overhead. DiLoCo naturally scales the learning rate accordingly. 
To test the synchronization overhead in a federated run, two clients on LUMI have started with different round synchronization intervals $H$. The per-round wall-clock time into compute (local training plus periodic held-out validation) and overhead (inter-site weight transfer plus checkpoint I/O, neither of which touches the GPU). Figure \ref{fig:barrier}(b) decomposes the overhead. At $H=1000$ synchronisation steps the training effectively takes 95\% of the total round time. 

\begin{table}[!h]
\centering
\caption{Per-round wall-clock time split into compute and synchronization
overhead, across local step counts $H$. All three runs use identical
transport (inline/gRPC) and sites (LUMI).}
\label{tab:overhead-vs-h}
\begin{tabular}{lrrrr}
\toprule
$H$ & Compute (s) & Overhead (s) & Total (s) & Overhead share \\
\midrule
100  & 235.9  & 111.9 & 347.8  & 32.2\% \\
1000 & 1774.2 & 107.0 & 1881.2 & 5.7\%  \\
2000 & 3601.8 & 115.8 & 3717.7 & 3.1\%  \\
\bottomrule
\end{tabular}
\end{table} Overhead is essentially flat in absolute terms across a 20$\times$ range of
$H$ (Table \ref{tab:overhead-vs-h}). Because it is sized by the model, not
by training length, communication alone varies by under 10\,s across all
three runs (94.5\,s / 90.4\,s / 98.7\,s). The overhead's share of wall-clock time falls
from 32.2\% at $H{=}100$ to 3.1\% at $H{=}2000$, as compute grows linearly. Finding a good $H$ must balance overhead costs and model convergence as higher synchronization steps will suffer more from training drift.

\textbf{Transport implementation: gRPC vs. HTTP.} At $H=100$, we compare two implementations of the weight-exchange step:
embedding weights directly in Flower's preferred gRPC control messages (\emph{inline}) versus
publishing them to a file for clients to fetch over HTTP (\emph{blob}).

\begin{table}[!h]
\centering
\caption{Per-round overhead (communication + checkpoint + evaluation) under
two transport implementations, at $H{=}100$.}
\label{tab:transport}
\begin{tabular}{lrrrr}
\toprule
Transport & Overhead (s) & Median (s) & $\Delta$ vs.\ gRPC \\
\midrule
inline / gRPC & 176.0 & 174.0 & --- \\
blob / HTTP   & 161.8 & 155.6 & $-14.2$\,s ($-8.1\%$) \\
\bottomrule
\end{tabular}
\end{table}

The blob variant reduced round overhead by 14.2\,s on average
(Table~\ref{tab:transport}). Flower's gRPC in our implementation serializes the
model as 310 separate per-tensor byte fields rather than one packed
stream, adding more overhead that scales with tensor count; the
blob path instead streams tensors into a single sequential file, and
publishes weights once
Thus, for LLM pretraining HTTP-based communication is preferred.

\begin{table}[!h]
\centering\small
\caption{The DARL coordinator's ledger for the three-site run (Frontier, LUMI, Snellius against one VM; 2,692 blocks of 1,024 windows). \emph{Lost} blocks of a vanished site holding them got returned to the pool on TTL expiry. Per-site committed counts sum to the coordinator's own total (2,045), so nothing was skipped either. No site was configured with a share: the 5.5 : 3.0 : 1.0 split is what the sites' own commit rates produced.}
\label{tab:darl}
\begin{tabular}{lrrrrrrrr}
\toprule
site & ranks & RTT & \# of entry & \multicolumn{4}{c}{blocks} & TTL \\
\cmidrule(lr){5-8}
 & & (ms) &   & grants & granted & committed & lost & expiries \\
\midrule
Frontier & 8 & 347 & 2 & 173 & 1,195 & 1,181 & 14 & 2 \\
LUMI & 8 & 70 & 1 & 96 & 656 & 649 & 7 & 1 \\
Snellius & 4 & 3 & 1 & 57 & 219 & 215 & 4 & 1 \\
\midrule
total & & & & 326 & 2,070 & 2,045 & 25 & 4 \\
\bottomrule
\end{tabular}
\end{table}

\textbf{Data leasing (Table~\ref{tab:darl}).} In the three-site run, Frontier, LUMI and Snellius leased from one VM over 23.8\,h, with per-site round trip times (RTT's) from 3\,ms to 347\,ms. The run ended with the sites committing 2,045 blocks. The per-site counts sum exactly to the coordinator's own total, so there was no duplication. Every site departure, crash or clean, cost exactly its one in-flight lease, and the TTL returned that lease to the pool. Frontier's first job died while it held nine blocks. The coordinator re-granted those nine blocks, five to LUMI and four to Frontier's own restart, and the sites successfully committed every one of them, with not duplicate blocks being committed.

The ledger makes two things quantitative. First, nobody configured a share, and no static share would have been right. Snellius delivered 32\% of an equal third of the corpus and Frontier delivered 173\% of it, a $5.5\times$ spread. A pre-split corpus would have turned that spread into an unfinishable slow shard. A crashed site would also have removed its whole shard from the epoch, rather than one lease. Every site runs the identical loader over the identical corpus, so all three walk the same prefix. The unique corpus trained then collapses to the largest site's share alone, 1,181 blocks against 2,045, which is $1.73\times$ less for the same compute.

Second, a departure forfeits exactly one lease. The coordinator grants one span at a time. A vanishing site can therefore lose only the $\lceil HBWA/K \rceil$ blocks it was holding. Long inner phases, $H$, make the failure unit more expensive. The TTL's $\alpha\,\bar t_{\text{step}}$ scaling keeps live sites with longer phases safe. Meaning that the effect of a crash is dependent on $H$.

\textbf{Scheduler.}\label{sec:scheduler}
The queue-aware scheduler is tested against a Qwen3-0.6B, 100B-token-target federated run across LUMI and Snellius, under a QoS of 4 nodes per site, a single node per allocation and a maximum walltime of 8h, comparing three queue scenarios. Table~\ref{tab:queue-scenarios-idle0} assumes 0\% barrier awaits for simplicity: neither site ever waits on the other. \emph{Wait} is each site's queue delay before its first job; \emph{Active} is how long it trains before the target is reached; \emph{Jobs} counts individual node-allocations (LUMI chains 1h jobs, Snellius chains 8h jobs, both back-to-back with no re-queueing gap) to cover \emph{Active}; \emph{Tokens} is that site's resulting share of the 100B target. Snellius' NVIDIA H100 are significantly faster in throughput than LUMI as can be seen in the number of processed tokens for Snellius, suggesting the throughput-awareness of the scheduler.

\begin{table}[!h]
\centering
\footnotesize
\caption{Scheduler allocation for a 100B-token Qwen3-0.6B run across LUMI and
Snellius under three queue conditions, assuming 0\% idle.}
\label{tab:queue-scenarios-idle0}
\begin{tabular}{@{}llrrrrr@{}}
\toprule
Scenario & Site & Shape (GPU$\times$h) & Wait & Active & Jobs & Tokens \\
\midrule
\multirow{2}{*}{A: No queue (1.11 d to 100B)}
  & LUMI     & 8$\times$1h & 0.1 h & 26.6 h & 108 & 29.9B  \\
  & Snellius & 4$\times$8h & 0.2 h & 26.5 h &  16 & 70.1B  \\
\addlinespace
\multirow{2}{*}{B: Snellius busy ~40h, LUMI free (2.28 d to 100B)}
  & LUMI     & 8$\times$1h &  0.2 h & 54.4 h & 220 & 61.3B \\
  & Snellius & 4$\times$8h & 40.0 h & 14.6 h &   8 & 38.7B \\
\addlinespace
\multirow{2}{*}{C: LUMI busy 40h, Snellius free (1.58 d to 100B)\textsuperscript{*}}
  & LUMI     & 8$\times$1h & 40.0 h &  0.0 h &   0 &  0.0B \\
  & Snellius & 4$\times$8h &  0.2 h & 37.8 h &  20 & 100.0B \\
\bottomrule
\end{tabular}

\vspace{4pt}
\raggedright\footnotesize\textsuperscript{*}Snellius alone reaches 100B (37.8h)
before LUMI's 40h wait is over, so LUMI never joins.
\end{table}

\section{Conclusion}
In this paper we have described a system for LLM pre-training across three heterogeneous supercomputers differing in hardware, network connectivity, and batch-scheduling policy. On top of DiLoCo, it introduces elastic aggregation, DARL, and queue-aware planning to provide the flexibility and resilience that batch-scheduled infrastructure requires. After 20{,}000 optimiser steps, the three-site run reaches a held-out perplexity of 34.7, against 28.2 for a centralised baseline. At $H=1000$ only 6\% measured communication overhead. Data coverage held throughout: across four site departures in a three-site run, 1.2\% of granted blocks were reclaimed and no block was lost or trained twice. The value of cross-site training today is therefore operational rather than competitive: it turns fragmented allocations into one training run, at an overhead we can quantify.

\textbf{Limitations}. This was a proof of concept at a single small model scale, one epoch, and one site topology, without downstream benchmarking. Future work should test production-scale training across more facilities, with stronger fault tolerance and a planner informed by historical queue traces rather than only live probes.

\FloatBarrier
\bibliographystyle{abbrvnat}
\bibliography{refs}

\newpage
\appendix
\section{HPC sites}\label{app:systems}
In Table \ref{tab:systems} the HPC sites for the experiments are detailed. All experiments were run under non-root privileges.

\begin{table}[!h]
\centering
\caption{The three systems in this work with their hardware, software, network connectivity and batch configurations.}
\small
\setlength{\tabcolsep}{4pt}
\renewcommand{\arraystretch}{1.15}
\begin{tabular}{@{}p{0.20\linewidth}p{0.24\linewidth}p{0.24\linewidth}p{0.24\linewidth}@{}}
\toprule
 & \textbf{Snellius} & \textbf{LUMI} & \textbf{Frontier} \\
\midrule
\multicolumn{4}{@{}l}{\emph{Hardware}} \\
Accelerator & $4\times$ NVIDIA H100 & $4\times$ AMD MI250X & $4\times$ AMD MI250X \\
Devices per node & 4 GPUs & 8 GCDs & 8 GCDs \\
Intra-node link & NVLink / NVSwitch & Infinity Fabric & Infinity Fabric \\
Inter-node fabric & InfiniBand NDR 200 
 & $4\times$ Slingshot-11 & $4\times$ Slingshot-11 \\
Parallel filesystem & GPFS
 & Lustre & Lustre \\
\addlinespace
\multicolumn{4}{@{}l}{\emph{Software environment}} \\
Stack & CUDA & ROCm & ROCm \\
Delivery & bare-metal modules & Apptainer container & Apptainer container \\
\addlinespace
\multicolumn{4}{@{}l}{\emph{Connectivity}} \\
Compute-node egress & direct outbound & outbound via NAT 
 & HTTP(S) forward proxy \\
Long-lived gRPC & supported & supported & reaped by proxy loadbalancer \\
\addlinespace
\multicolumn{4}{@{}l}{\emph{Slurm batch system}} \\
Slurm REST API & exposed
 & not exposed & selectively exposed \\
Cron / timed jobs & unavailable & unavailable & unavailable \\
Collector mechanism & chained Slurm jobs &  chained Slurm jobs & user cronjob \\
Sub-node allocation & yes (CPU partition) & yes (\texttt{small} partition) & no (full node only) \\
Max walltime & 120\,h 
 & 48\,h (\texttt{standard} partition) 
 & 24\,h 
 \\
\bottomrule
\end{tabular}
\label{tab:systems}
\end{table}

\section{Training configuration}\label{app:setup}
Below in Table \ref{tab:setup} is the language model setup detailed. 
\begin{table}[h]
\centering
\small
\begin{tabular}{ll}
\toprule
Model               & Qwen3-0.6B, 28 layers \\
Tokeniser           & OpenEuroLLM 128k (131{,}328-row tied embedding) \\
Checkpoint          & 1.32\,GiB\\
Data                & C4, 5B tokens \\
Sequence length     & 2048 \\
Precision           & bfloat16 \\
Peak LR             & $3\times10^{-4}$ \\
Micro batch size    & 8 \\
Global batch size   & 96 by accumulation, or 32 Snellius $+$ 64 LUMI/Frontier \\
\bottomrule
\end{tabular}
\caption{Training configuration shared by all runs.}
\label{tab:setup}
\end{table}

\section{License}
All external assets are used under their published licences: torchtitan~\citep{torchtitan} (BSD-3-Clause, \copyright{} Meta Platforms), which also supplies the Qwen3~\citep{qwen3} model definition; Flower~\citep{flower} 1.33 (Apache-2.0); the OpenEuroLLM 128k tokeniser (Apache-2.0); and the C4 corpus~\citep{c4} (ODC-BY, subject to the Common Crawl terms of use). The code released with this paper is Apache-2.0.

\end{document}